# A macroscopic test of the Aharonov-Bohm effect.

Adam Caprez, Brett Barwick, and Herman Batelaan*

*Department of Physics and Astronomy, University of Nebraska-Lincoln, Lincoln NE 68588, USA*

*email: hbatelaan2@unl.edu

The Aharonov-Bohm (AB) effect is a purely quantum mechanical effect. The original (classified as Type-I) AB-phase shift exists in experimental conditions where the electromagnetic fields and forces are zero. It is the absence of forces that makes the AB-effect entirely quantum mechanical. Although the AB-phase shift has been demonstrated unambiguously, the absence of forces in Type-I AB-effects has never been shown. Here, we report the observation of the absence of time delays associated with forces of the magnitude needed to explain the AB-phase shift for a macroscopic system.

In 1918, Weyl proposed a scaling of space-time in an attempt to combine electromagnetism and general relativity with what he coined a gauge theory [1]. Einstein subsequently pointed out that Weyl's space-time scaling led to contradictions [2]. Yang overcame Einstein's objections by modifying Weyl's idea to use a phase change instead of a gauge change [3]. Aharonov and Bohm (AB) proposed the use of magnetic flux enclosed in an electron interferometer to detect the phase shift [4]. Chambers, following the proposal, demonstrated the now famous AB-effect [5]. Later efforts [6-9] include Tonomura's beautiful experiments using magnetic toroids. Excellent agreement was found between the measured phase shift and the theoretical prediction:

$$\Delta\varphi = \frac{q}{\hbar}\oint \vec{A}\cdot d\vec{l},\qquad(1)$$

where the charge of the electron is $q$, the vector potential is $\vec{A}$, and the contour encloses the magnetic flux of a solenoid.

The original magnetic and electric AB-effects (Type-I) are distinguished from the neutron scalar AB-effect and the Aharonov-Casher effect (Type-II) by the absence of any electromagnetic fields [10]. The original AB-effect shows the non-local effect of electromagnetic fields [4]; electron wave packets are influenced although they travel in field-free regions of space. However, the potentials must not be zero in the same region to obtain a non-zero phase shift. As a consequence, it is often stated that the status of potentials is elevated to that of physically relevant entities, instead of just mathematical tools as they are in classical electromagnetism [11]. The demand of local gauge invariance, associated with the use of potentials in quantum mechanics, has been a guiding principle in the construction of modern field theory [12].

Given the pivotal importance of Type-I AB-effects, the current experimental situation is surprising. The absence of forces for the magnetic AB-effect has never been experimentally verified, while the electric AB-effect has escaped detection altogether [13]. The former is the topic of investigation in this letter.

To establish the absence of a force, an experimental signature is needed. In semi-classical theory, an unknown force could shift the electron wave packet by an amount $\Delta y$. This would introduce an electron wave phase shift [14],

$$\Delta \varphi = p_y \Delta y / \hbar ,  \qquad (2).$$

where $p_y$ is the electron's momentum. The predicted and experimentally confirmed value of the AB-phase shift would only be matched if

$$\frac{1}{\hbar} p_y \Delta y = \frac{q}{\hbar} \oint \vec{A} \cdot d\vec{l} = \frac{q}{\hbar} B_0 A . \qquad (3)$$

The difference between such a semiclassical theory and the generally accepted theory is that for the first one the electron wave packet shifts, while for the second the wavepacket is multiplied by the AB-phase factor and does not shift.

Thus, a crucial experiment has to be performed that demonstrates both the phase shift and absence of forces for the same experimental situation. The need for such a test was pointed out more than two decades ago by Zeilinger [15], but has not been done for Type-I effects. Such an experiment is referred to as testing the dispersionless nature of the AB-effect [16]. Dispersionless is defined as the independence of the phase shift magnitude with electron velocity. The result of a dispersionless interaction is that the average position of the electron wavepacket propagates as if in free space, or in other words in the absence of a force.

Using a non-zero displacement $\Delta y$ that satisfies Eq. (3) introduces a time delay (for small electron velocity changes) of

$$\Delta t = \Delta y / v_0 = B_0 A q / (m v_0^2) . \qquad (4)$$

Thus another experimental signature that can establish the absence or presence of a force which could potentially explain the AB-effect is a time delay measurement (Fig. 2).

It may appear exceedingly unlikely that a force of the correct magnitude would be present in the description of this physical system. However, in 2002, a force expression was calculated that yields Eq. (3)[14].

To arrive at Eq. (3) via a force approach, consider a solenoid as a stack of current loops. The solenoid axis is chosen along z. The magnetic field for a point charge at position $(x_e, y_e, z_e)$ moving non-relativistically past the solenoid (located at the origin) with constant velocity $\vec{v} = v_0 \hat{j}$ is given by

$$B(\vec{r},t) = \frac{v_0 q \mu_0}{4\pi} \frac{((z-z_e)\hat{i} - (x-x_e)\hat{k})}{((x-x_e)^2 + (y-y_e)^2 + (z-z_e)^2)^{3/2}}. \tag{5}$$

Charge carriers are flowing in each current loop, and Lorentz forces due to the electron act on them:

$$\vec{F} = \int I(d\vec{l} \times \vec{B}) =$$

$$\int I\hat{\phi} \times \frac{v_0 q \mu_0}{4\pi} \frac{((z-z_e)\hat{i} - (x-x_e)\hat{k})}{\left((x-x_e)^2 + (y-y_e)^2 + (z-z_e)^2\right)^{\frac{3}{2}}} r d\phi. \tag{6}$$

Performing the integration and looking at the y-component yields the force,

$$F_y = \frac{B_0 A v_0 q}{4\pi} \frac{4(x_e)(y_e)}{\left((x_e)^2 + (y_e)^2\right)^2}. \tag{7}$$

Boyer arrives at this force expression by modeling the solenoid as an infinite line of magnetic dipoles [14]. In our experiment the solenoid has a high permeability ($k$) iron core. This core can be modeled with magnetic dipoles aligned by the solenoid field. Boyer's derivation can thus be applied to the core. As a result the force of the electron on the solenoid-core system increases by a factor $k \sim 150$.

The change in velocity due to the force is given by $\Delta v_y^{(+)}(t) = \frac{1}{m}\int_{-\infty}^{t'=t} F_y(t')dt'$ and the change in pathlength is $\Delta y^{(+)}(t) = \int_{-\infty}^{\infty} \Delta v_y^{(+)}(t)dt$, where the "+" indicates that the electron passes on the $x > 0$ side of the solenoid. Consequently, the difference in pathlength for electron paths passing on either side of the solenoid is given by $\Delta Y = \Delta y^{(+)} - \Delta y^{(-)}$. This physical path length difference leads to a semi-classical phase shift of

$$\Delta \phi = \frac{p_y}{\hbar} \Delta Y = \frac{m v_0}{\hbar} \frac{B_0 A q}{m v_0} = \frac{B_0 A q}{\hbar}, \tag{8}$$

matching the AB-phaseshift. This intriguing result further motivates an experimental test, but we emphasize that regardless of the validity of any particular force approach, a delay time measurement can rule out the class of all semi-classical "force" theories yielding Eq. (4). We also note that the generally accepted theoretical understanding is that such terms as hidden momentum compensate any force on the solenoid, so that neither the solenoid nor the passing electron are expected to experience a force [17-19].

In this letter, a time-of-flight experiment for a macroscopic solenoid is performed. To start this time-of-flight experiment, a femtosecond laser pulse is used to extract electrons from a field emission tip [20]. The electron pulse then passes between two identical solenoids. The two solenoids are connected through high permeability magnet iron bars, and form a square magnetic toroid (Fig. 2). This arrangement reduces magnetic flux leakage. A magnetic cylindrical shield can also be placed between the solenoids along the electron's propagation direction to shield unwanted magnetic fields. Finally, the arrival of the electron is detected with a channelplate and a time-of-flight spectrum is obtained.

The primary result is that as a function of the current through the solenoids no time delay is observed (Fig. 3), thus signaling the absence of forces. The inset in Fig. 3 shows examples of a time-of-flight spectrum for three different solenoid currents. The scatter of the arrival times is ±0.1 ns.

For the applied current $I$, the 2.5 mm diameter solenoid gives a magnetic flux of $B_0 A = k\mu_0 InA$, where $k \sim 150$ is the magnetic permeability of the iron core, $\mu_0$ the permeability of free space, and $n=3$/mm is the winding density. For these parameters the time delay (Eq. (4)) is indicated in Fig. 3 by the solid line.

Changes in the magnetic flux are registered continuously during the experiment with a magnetic induction pick-up coil. The magnitude of the flux deduced from the induction spikes agrees with that obtained from the measured solenoid current.

Although the measured time delays are zero (Fig. 3), the experimental arrangement is sensitive at the nanosecond scale to electromagnetic forces. To demonstrate this, time delay data for different acceleration voltages are given in Fig. 4. The arrival time as a function of electron energy agrees well with a classical ballistic model. The fact that a classical description is sufficient further motivates the term "macroscopic."

Removing the magnetic shielding in the experiment does not change the outcome (red circles in Fig. 3). However, the electric field is expected to be shielded by the solenoid [21], which acts as a Faraday cage. In some experiments including ours, a metal shield contains the solenoid. It is also well established that for interaction times as fast as at least $10^{-14}$ s the charge carriers in a metallic surface have sufficient time to form an image charge and setup the shielding electric field. This timescale is well established and related to the inverse plasmon frequency [22]. For the specific case of a quantum mechanical electron wave experiment, we observed the effect of the image charge on an electron diffraction pattern for a 100 nm gold coated grating [23]. It leaves no doubt that the charge carriers of a metal shield in larger structures, which includes all AB-experiments to date, shield the interior of a solenoid. On the one hand the image charge interaction with the electron leads to time delays.

It is straightforward to show that these timedelays are negligible compared to those of Eq. 4, as well as current independent. On the other hand, the moving image charge suggests that dynamics, which are excluded from the generally accepted explanations [17-19], could be important. In this sense our experiment is not merely a demonstration, but settles the question whether there is a force or not.

In conclusion, we have shown that there is no force acting on an electron passing by a "macroscopic" solenoid of a magnitude that can potentially explain the AB-effect. All force explanations leading to a time delay (Eq. (4)) can be ruled out.

**Acknowledgements:** This material is based upon work supported by the National Science Foundation under Grant No. 0653182.

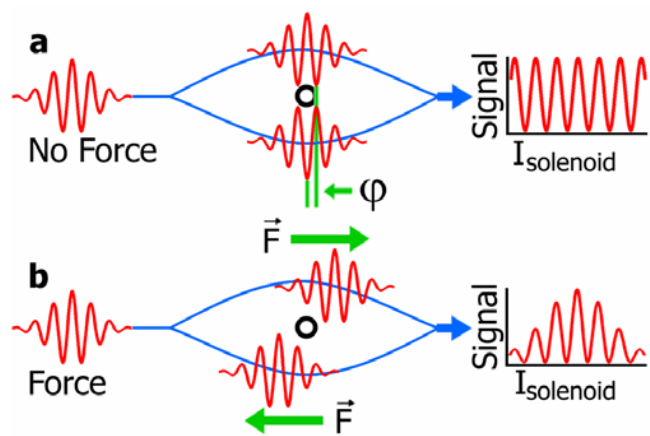

FIG. 1. Force versus phase. Electrons can follow two paths in an electron interferometer (blue lines) enclosing a solenoid (black circle). The Aharonov-Bohm interference fringes (Signal) are thought to be due to a multiplicative phase factor (a). A force can shift the electron wave packet and cause similar interference fringes (b). The difference is that for a force the fringes do not extend beyond the electron coherence length.

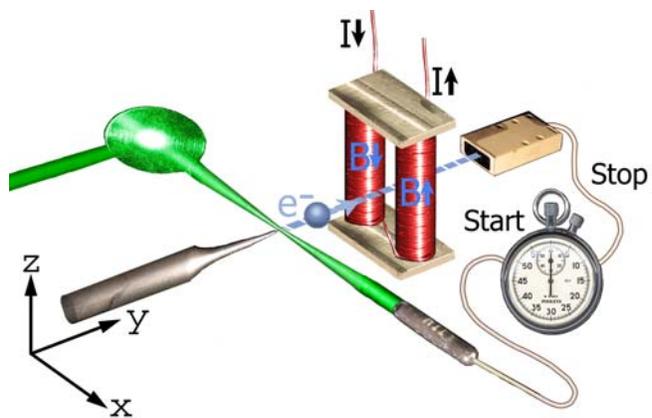

FIG 2. Experimental schematic. A laser induced electron pulse passes between solenoids. A time delay measurement establishes the presence or absence of forces on the electron.

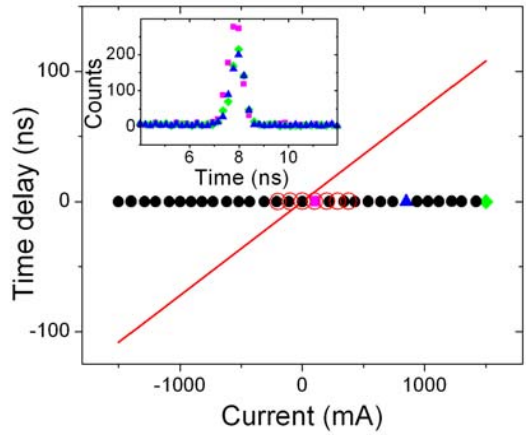

FIG 3. Delay times. No delay for 40 eV electrons is observed as a function of solenoidal current, signaling the absence of a force consistent with the Aharonov-Bohm prediction. A force necessary to explain the AB-effect would produce delays indicated with the red solid line (Eq. (4)). The square, the triangle and the diamond data point each correspond to a time-of-flight spectrum (inset).

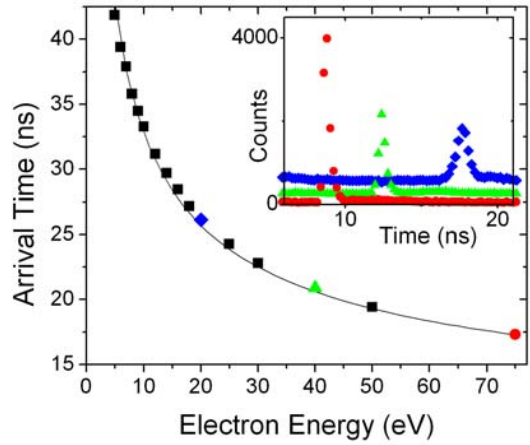

FIG 4. Energy-time delays. Electron arrival times as a function of energy follow a ballistic behavior. The diamond, the triangle and the square data point each correspond to a time-of-flight spectrum (inset). The time delay sensitivity is sufficient to detect the presence of forces necessary to explain the AB-effect.